\documentclass[aps %showpacs,
preprintnumbers,amsmath,amssymb,geometry]{revtex4}
\usepackage{graphicx}
\usepackage{dcolumn}% Align table columns on decimal point
\usepackage{bm}% bold math

\usepackage{graphicx}
\usepackage{epsfig}
\usepackage{color}
\usepackage[T1]{fontenc}
\usepackage[utf8]{inputenc}
\usepackage{combelow}% provides \cb to place comma below character

\usepackage{newunicodechar}
\newunicodechar{ș}{\cb{s}}
\newunicodechar{ț}{\cb{t}}

\usepackage{amssymb,amsfonts,amsmath}

\newcommand{\ms}{\medskip}

\begin{document} %.........................................................
 \title{Liouville soliton surfaces obtained using Darboux transformations}

 \author{S. C.  Mancas}
\email{mancass@erau.edu, http://orcid.org/0000-0003-1175-6869}
\affiliation{Department of Mathematics, Embry--Riddle Aeronautical University,\\ Daytona Beach, FL. 32114-3900, USA}

 \author{K. R. Acharya}
 {\email{acharyak@erau.edu, http://orcid.org/0000-0003-3551-7141}
 \affiliation{Department of Mathematics, Embry--Riddle Aeronautical University,\\ Daytona Beach, FL. 32114-3900, USA}

 \author{H. C. Rosu}
 {\email{hcr@ipicyt.edu.mx, http://orcid.org/0000-0001-5909-1945}
  \affiliation{IPICyT, Instituto Potosino de Investigacion Cientifica y Tecnologica,\\
  Camino a la presa San Jos\'e 2055, Col. Lomas 4a Secci\'on, 78216 San Luis Potos\'{\i}, S.L.P., Mexico}
%\pacs{02.30.Ik, 02.30.Hq, 03.65.Fd}

%%%%%%%%%%%%%%%%%%%%%%%%%%%%%%%  AAA  %%%%%%%%%%%%%%%%%%%%%%%%%%
\begin{abstract}
\noindent %
In this paper, Liouville soliton surfaces based on some soliton solutions of the Liouville equation are constructed and displayed graphically, including some of those corresponding to Darboux-transformed counterparts. We find that the Liouville soliton surfaces are centroaffine surfaces of Tzitzeica type and their centroaffine invariant can be expressed in terms of the Hamiltonian. The traveling wave solutions to Liouville equation from which these soliton surfaces stem are also obtained through a modified variation of parameters method which is shown to lead to elliptic functions solution method.
\end{abstract}
%%%%%%%%%%%%%%%%%%%%%%%%%%%

\centerline{Physica Scripta 98, 075227 (2023)}

\maketitle
%1111111111111111111111111111111111111111
\section{Introduction} \label{sec1}

In this paper, we introduce soliton surfaces based on the Liouville sech$^2$ soliton solution and also of its Darboux/auto-B\"acklund transformations. The employed Liouville soliton is one of the common solutions obtained by using linear arbitrary functions $\Phi(u)$ and $\Psi(v)$ in the Liouville representation of the general solution. Although the Darboux/auto-B\"acklund transformations are known for more than a century in mathematics, and the concept of soliton surface has been introduced in mathematical physics in the 1980s \cite{sym83,sym85}, we could not find any discussion of the Liouville soliton surfaces in the recent literature, e.g., \cite{cies98,gp12,gpr14}. This gap in the literature served us as a motivation for writing this paper. Another motivation is that soliton surfaces are geometrical analogs of gauge theories of (super)strings, spin systems, and chiral integrable models of elementary particles in which the interaction is not generated by considering interaction Lagrangians, but has pure geometric origin related to the curvature of the soliton surface and other geometric and topological concepts \cite{sym83,sym85,perel87,dt07}.

\ms

This paper is structured as follows. In Section 2, some very basic results about the Liouville equation with some historical flavor are provided.
In Section 3, the affine geometric representation is described following methods originally due to Tzitzeica. It is shown that the Liouville soliton surfaces are centroaffine surfaces and their centroaffine invariant is the Hamiltonian. In Section 4, a modified variation of parameters method is shown to be of usage in obtaining the same Liouville soliton solution providing also a direct way to the Hamiltonian of the system. In section 5, the Lax pair of matrices introduced by Mikha\u{\i}lov \cite{Mikh} are used to obtain the coordinates of the soliton surfaces in the case of Darboux-transformed solutions and two such surfaces obtained in this way are displayed graphically. A concluding section summarizes the results and presents some perspectives for possible future work.

\ms

\section{The Liouville equation: A brief historical sketch}

The Liouville nonlinear equation first occurred around 1850 as a result obtained by Liouville when he carefully studied the surfaces of constant curvature, cf. one of his notes, in the fifth edition of the famous textbook by Monge \cite{mo} curated by Liouville. He  was led to the following partial differential equation
%% eq. 1
\begin{equation}\label{eq2}
{(\log \lambda )}_{uv}\pm\frac{\lambda}{2a^2}=0~,
\end{equation}
where $a$ is a real arbitrary constant.
%Our goal is to first obtain the traveling wave solution to this equation then derive the geometrical representation of such solutions.
By letting $\alpha=\mp\frac{1}{2a^2}$, and denoting $\Lambda= \log \lambda$, this equation may be written in the equivalent form
%% eq. 2
\begin{equation}\label{eq1}
E(\Lambda)\equiv \Lambda_{uv}-\alpha e^{\Lambda}=0~.
\end{equation}

In his notes, Liouville  presented a very simple method to the complete integral of \eqref{eq2} through some geometric considerations related to the properties of the sphere \cite{mo,Lio}. His solution, involving two arbitrary functions $\Phi(u)$ and $\Psi(v)$ was written as
%% eq. 3
\begin{equation}\label{eq3}
\lambda(u,v)=\frac{2}{\mp\alpha}\frac{\Phi_u\Psi_v~e^{\Phi(u)+\Psi(v)}}{[1\pm e^{\Phi(u)+\Psi(v)}]^2}~.
\end{equation}
%where $'$ indicate ordinary derivatives with respect to their variables.
One can easily see that \eqref{eq3} can be also written as
%% eq. 4
\begin{equation}\label{eq3a}
	\begin{array}{l}
\lambda(u,v)=\mp\frac{1}{2\alpha}\Phi_u\Psi_v {\rm sech}^2\left(\frac{\Phi(u)+\Psi(v)}{2}\right)~ \quad {\rm or} \\
\lambda(u,v)=\pm\frac{1}{2\alpha}\Phi_u\Psi_v{\rm cosech}^2\left(\frac{\Phi(u)+\Psi(v)}{2}\right)~,
\end{array}
\end{equation}
corresponding to bound and singular at the origin solutions, respectively. An even simpler form of \eqref{eq3} can be obtained
by using the substitutions $\log \phi =\Phi$, and  $\log \psi =\Psi$,
%%....eq. 5
\begin{equation}\label{eq3b}
\lambda(u,v)=\frac{2}{\mp\alpha}\frac {\phi_u\psi_v}{\left(1\pm \phi\psi\right)^2}~.
\end{equation}
The Liouville equation stayed nearly dormant for more than a century until its application as a key dynamical equation of motion for the string degrees of freedom in quantum field theory, two-dimensional gravity, and soliton particle physics models has been revealed during the decade 1980. We recommend the reader
the reviews \cite{bn, teschner}. Interestingly, it has been found that some singular Liouville solutions may play a role in the relativistic string dynamics since they do no make the string singular and consequently their mathematics has been further developed in \cite{dpp,jkm}.

\section{An affine surface and the B\"acklund Transformation}
It was noted by the Romanian mathematician Gheorghe \cb{T}i\cb{t}eica,  a.k.a. Georges Tzitz\'eica  \cite{TG} who studied under Darboux and Goursat, that in some asymptotic coordinates $u,v$, surfaces $\Sigma$ parametrized by $s^1(u,v), ~s^2(u,v), ~s^3(u,v)$ have the property that the  Gaussian  curvature  ${\cal K}$ is proportional to the fourth power of the distance $D$ from the origin $O$ to the tangent plane to the surface $\Sigma$  at some point $P$ of coordinates $(s^1,s^2,s^3)$. Thus, Tzitz\'eica introduced the centroaffine  constant invariant ${\cal I}$ defined by
%%%%....eq. 6
\begin{equation}\label{inv}
{\cal I} =\frac{{\cal K}}{D^4}~.
\end{equation}

The idea to construct surfaces in this way was also used by Tzitz\'eica for the equation rediscovered by Dodd and Bullough (the Dodd-Bullough equation),  and further he introduced a mapping (auto-B\"acklund transformation) between surfaces in his papers \cite{TG1,TG2}, and book published in French \cite{TG}.

These surfaces therefore admit the  affine representation given by the coordinates $(s^1, s^2, s^3)$  which are three linearly independent integrals of the linear system
%% system 7
\begin{align} \label{eq3bis}
 \begin{array}{ll} &s_{uu}=a s_{u}+b s_{v}~, \\
& s_{uv} =\alpha \lambda s~, \\
  &s_{vv}=a' s_{u}+b' s_{v}~.
   \end{array}
   \end{align}
For the specific values of $a=\frac{\lambda_u}{\lambda}$, $b=a'=0$,  and $b'=\frac{\lambda_v}{\lambda}$, {\em any} solution $\lambda(u,v)$ of the Liouville equation represents the compatibility solution of the system
%% system 8
   \begin{align} \label{eq4bis}
 \begin{array}{ll} & s_{uu}= \frac{\lambda_u}{\lambda} s_{u}~, \\
& s_{uv} =\alpha \lambda s~,\\
  &s_{vv}=\frac{\lambda_v}{\lambda}s_{v}~.
   \end{array}
   \end{align}

%%%%

The first  Lax pair for \eqref{eq1}, known by Tzitz\'eica,  is given by
%% Lax eqs 9
\begin{equation}\label{sys3}
\begin{array}{l}
S_u=LS\\
S_v=AS
\end{array} ~,
\end{equation}
where   $S(u,v)$ is a vector valued wave function given by $S^T=(s_v, s_u, s)$,
%$S = \begin{pmatrix} s_v \\ s_u \\ s\end{pmatrix}$,
and  $L$, $A$ are the third order matrices
 \begin{equation} \label{laz}
 L = \begin{pmatrix} 0 & 0 & \alpha e^\Lambda\\ 0 & \Lambda_u & 0\\ 0 &1 & 0 \end{pmatrix}, \,\,\,\,\,\,\, A  = \begin{pmatrix} \Lambda_v& 0 & 0 \\ 0 & 0 &  \alpha e^\Lambda\\1 &0 & 0 \end{pmatrix}. \end{equation}
 This pair satisfies  the compatibility condition $S_{uv}=S_{vu}$ which leads to
 %% eq. 11
 \begin{equation}
 L_v-A_u+[L,A]=\begin{pmatrix} -E(\Lambda) & 0 & 0\\ 0 & E(\Lambda)  & 0\\ 0 &0 & 0 \end{pmatrix}.
 \end{equation}

We are interested in generating solitonic surfaces corresponding to \eqref{eq2} and first Lax pair \eqref{laz} for which we use a linear form of  the general  functions given by
%% eq. 12
\begin{equation}\label{sys33}
\begin{array}{l}
\Phi(u)=2\sqrt{\alpha}~ku~, \\
\Psi(v)=-2\sqrt{\alpha}~\omega v~.
\end{array}
\end{equation}
Since we deal with solitons, we also use henceforth the right- and left-moving traveling coordinates, $\xi=ku-\omega v$ and $\zeta=ku+\omega v$, respectively, and also known as light-cone coordinates in the context of linear wave equations.
Thus, for $\alpha>0$, the soliton solution of \eqref{eq2} in the variable $\xi$ has the form, %=ku-\omega v$,
%% eq. 13
\begin{equation}\label{eq11}
\lambda(u,v)= 2 k\omega ~\mathrm{sech}^2\left(\sqrt{\alpha}\xi\right)~.
\end{equation}
This will be the seed solution used to generate $\Sigma$. To find the surface corresponding to these solutions, we  need to solve the linear system \eqref{eq4bis}.  By two quadratures the third equation leads to
%% eq. 14
 \begin{equation}\label{eqr}
s(u,v)=-c_1(u)\frac{\tanh \left(\sqrt{\alpha}\xi\right)}{\sqrt{\alpha} \omega}+c_2(u)=
\left[-\frac{c_1(u)}{\sqrt{\alpha}\omega}\sinh \left(\sqrt{\alpha}\xi\right)+c_2(u)\cosh \left(\sqrt{\alpha}\xi\right)\right]\text{sech}\left(\sqrt{\alpha}\xi\right)~,
 \end{equation}
while the second equation is satisfied for
%% eq. 15
\begin{equation}\label{eqg}
c_2(u)=\frac{c_1(u)'}{ 2 k \alpha \omega}~.
\end{equation}
%for which  (\ref{eqr}) is
% \begin{equation}\label{eqs}
% s(u,v)=\frac{1}{\alpha}\left[f'(u)-2k \sqrt{\alpha}f(u) \tanh \sqrt{\alpha}(k u- \omega v)\right].
% \end{equation}
To find $c_1(u)$ we substitute the last two in the first equation of (\ref{eq4bis}), which leads  to the third order linear differential equation
%% eq. 16
  \begin{equation} \label{eq21}
  c_1(u)''' - 4 \alpha k^2  c_1(u)'=  0~.
  \end{equation}
Thus, the  surface $\Sigma$ is  determined by the algebraic curves of the  cubic
%% eq. 17
\begin{equation}\label{curves}
m^3- 4 \alpha k^2 m=0
\end{equation}
with roots $m_1=0$, $m_{2,3}=\pm 2\sqrt{\alpha}k$ which lead to the three solutions
%% eq. 18
\begin{align} \label{card3}
 \begin{array}{ll}
  &c_1^1(u)= \sqrt{\alpha} \omega~,\\
 & c_1^2(u)= \sqrt{\alpha} \omega \sinh (2\sqrt{\alpha}k~u)~,\\
 & c_1^3(u) = \sqrt{\alpha} \omega \cosh (2\sqrt{\alpha}k~u)~.
\end{array}
 \end{align}

Using these functions in \eqref{eqg} and \eqref{eqr},  the three linearly independent  solutions of system (\ref{eq4bis})
%% eq. 19
\begin{align} \label{card4}
\begin{array}{ll}
&s^1(u,v)= \mathrm{sinh} \left (\sqrt{\alpha}\xi\right)\mathrm{sech}  \left(\sqrt{\alpha}\xi\right)~,\\
&s^2(u,v)= \cosh  \left (\sqrt{\alpha}\zeta\right)~\mathrm{sech}  \left ( \sqrt{\alpha } \xi\right)~, \\
&s^3(u,v)=    \mathrm{sinh}  \left ( \sqrt{\alpha} \zeta\right)~\mathrm{sech}  \left ( \sqrt{\alpha}\xi\right)~.\\
\end{array}
\end{align}

The Darboux-transformed Liouville solutions can be successively  generated  using the invariant transformations \cite{Bre}, \cite{Schief}
%% eq. 20
\begin{equation}\label{np2}
	\tilde \lambda^i=-\lambda +\frac{2}{\alpha}(\log s^i)_u(\log s^i)_v~,
\end{equation}
where $\lambda$ is the seed solution given by (\ref{eq11}),  and $s^i$ with $i=1,2,3$,  is any solution of \eqref{card4}. This type of transformation known as B\"acklund  transformation  represents a relation between two sets of independent solutions  $\lambda$ and $\tilde \lambda$ of the same PDE.

Using successively these three solutions, the new potentials $\tilde{\lambda}$ are
%% eq. 21
 \begin{align}
		\begin{array}{ll}
			&\tilde{\lambda}^1(u,v)= -2k\omega~ \mathrm{csch}^2 \left(\sqrt{\alpha}\xi\right)~,\\
			&\tilde{\lambda}^2(u,v)=-2k\omega~ \mathrm{sech}^2 \left(\sqrt{\alpha} \zeta \right)~, \\
			&\tilde{\lambda}^3(u,v)= 2k\omega ~\mathrm{csch}^2 \left(\sqrt{\alpha} \zeta\right)~.\\
		\end{array}
\end{align}
These  are also solutions to the  Liouville equation \eqref{eq2}, and can be infinitely many times generated using \eqref{np2}.

The  three solutions $s^i(u,v)$  allow us to generate a  parametric surface $\Sigma(s^1,s^2,s^3)$ in  $\mathbb{R}^3$  known as the  Liouville soliton surface.  To classify the surface,  notice the quadratic relation between these solutions defining the one-sheeted hyperboloid
  %% eq. 22
\begin{equation}\label{hyp}
\Sigma(u,v)\equiv[s^1(u,v)]^2+[s^2(u,v)]^2-[s^3(u,v)]^2=\frac{4k^2}{\alpha}~,
\end{equation}
 for which
 %% eq. 23
 \begin{equation} \label{zop}
 s^3=h(s^1,s^2)=\sqrt{[s^1(u,v)]^2+[s^2(u,v)]^2-\frac{4k^2}{\alpha}}~.
 \end{equation}

To calculate the centroaffine invariant for this surface, we use the curvature
%% eq. 24
\begin{equation}\label{in1}
{\cal K}=\frac{|h_{s^1s^1}h_{s^2s^2}-(h_{s^1s^2})^2|}{\left[1+(h_{s^1})^2+(h_{s^2})^2\right]^2}~,
\end{equation}
and distance $D$
%% eq. 25
\begin{equation}\label{dee}
D=\frac{|s^1h_{s^1}+s^2h_{s^2}-h(s^1,s^2)|}{\left[1+(h_{s^1})^2+(h_{s^2})^2\right]^{1/2}}~.
\end{equation}
For $\Sigma$,  we have then the invariant is
%% eq. 26
\begin{equation}\label{dr}
{\cal I}=\frac{\alpha^3}{(2k)^6}~.
\end{equation}
This indeed shows that the Gaussian curvature at the point $P$ on $\Sigma$ is proportional to the fourth power of the distance from the origin to the tangent plane to $\Sigma$ at $P$. In Fig.~{\ref{figure2},  we plot the Liouville surface $\Sigma$ for the seed sech square soliton solution for some fixed values of the set $k, \omega$, and $\alpha$.
%................. FIGURE 1
\begin{figure}[ht!]
	\centering
	\includegraphics[width=0.4\textwidth]{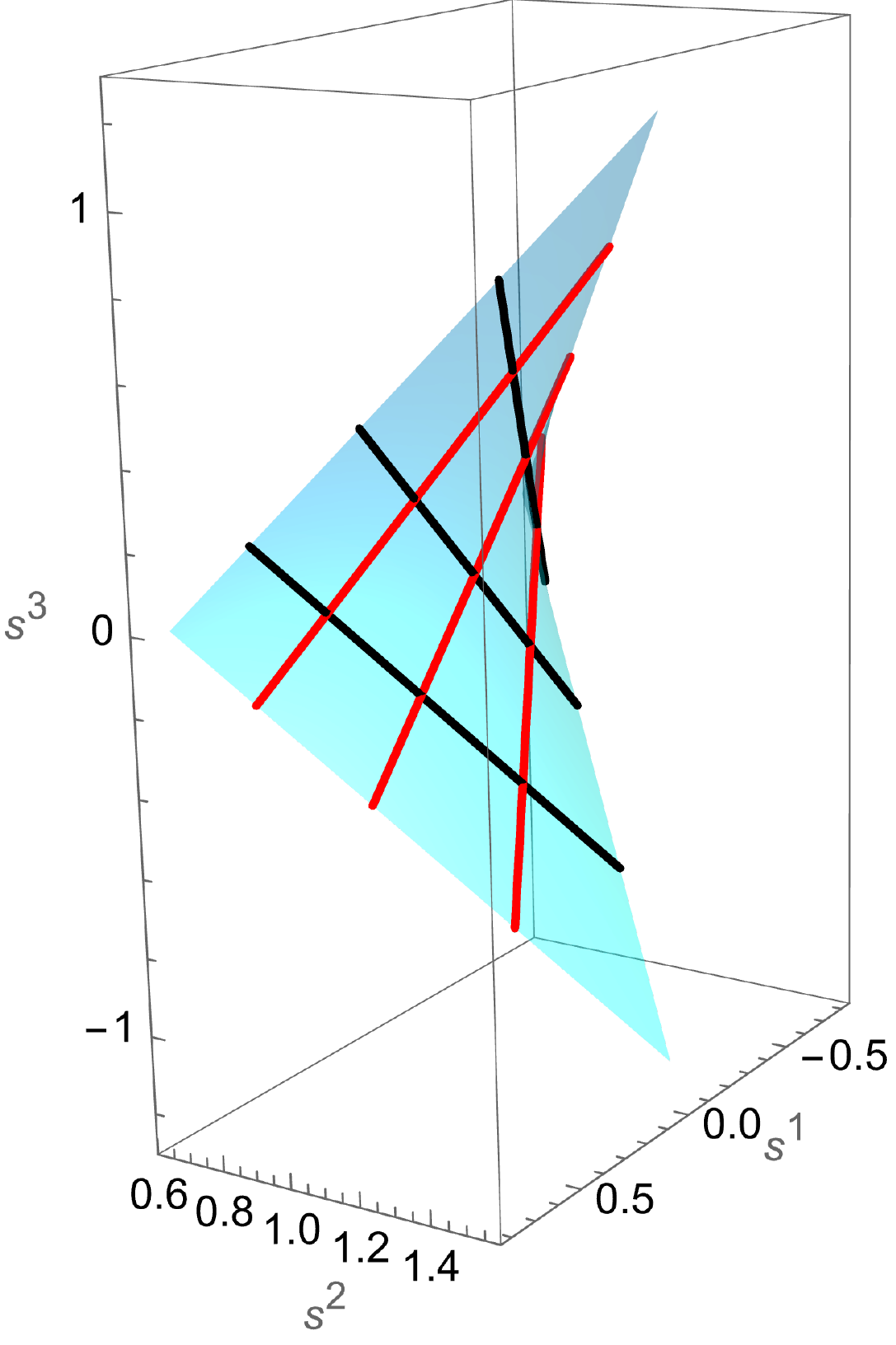}
	\caption{Using \eqref{zop}, we plot one octant of the Liouville soliton surface $\Sigma$.  This surface is a hyperboloid of one sheet for the solitary wave solution \eqref{eq11} with $k=1, \omega=1, \alpha=1/4$.
The centroaffine invariant is ${\cal I}=1/4^6$. The red curves are found  using $ v=-1/2,0,1/2$, while the black curves correspond to $u=-1/2,0,1/2$. All of these curves are lines,  since the one-sheet hyperbolioid is a doubly ruled surface.}
	\label{figure2}
\end{figure}
%..................

One can generalize these type of surfaces  by choosing any particular functions of interest, such as  general power-functions  that depend on some parameter $p$.  One example, given by  $\Phi(u)=2\sqrt{\alpha}~ku^p$ and $\Psi(v)=-2\sqrt{\alpha}~\omega v^p$ leads to the parametric soliton solution
%% eq. 27
\begin{equation}\label{eq11a}
\lambda(u,v;p)= 2 p^2\,k\omega\,(uv)^{p-1} ~\mathrm{sech}^2\left(\sqrt{\alpha}\xi^p\right)~,  %(ku^p-\omega v^p)\right]~.
\end{equation}
where $\xi^p=ku^p-\omega v^p$. The soliton solutions corresponding to $p=1,2,3$  are presented in  Figs.~{\ref{figa}} and {\ref{figabs}}.
%...................................FIGURE 2
\begin{figure}[ht!]
\centering
\includegraphics[width=0.495\textwidth]{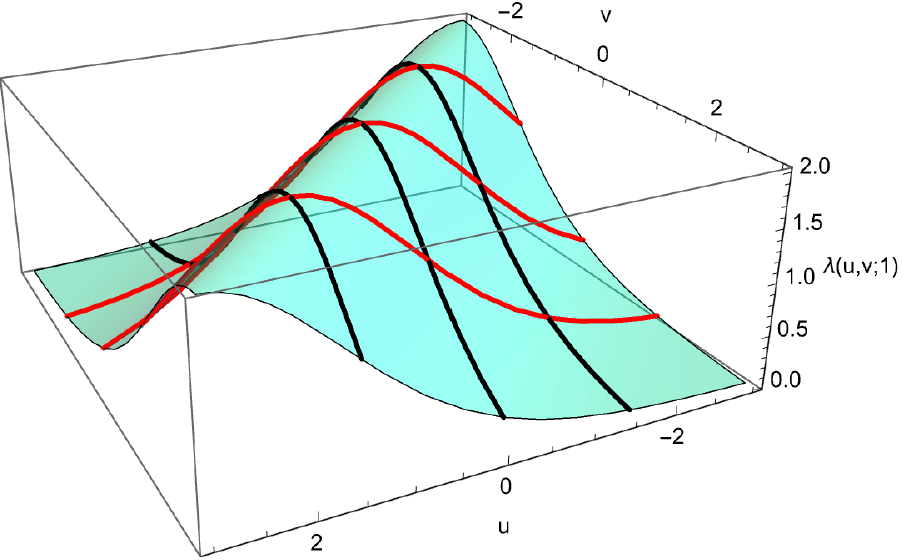}
\includegraphics[width=0.495\textwidth]{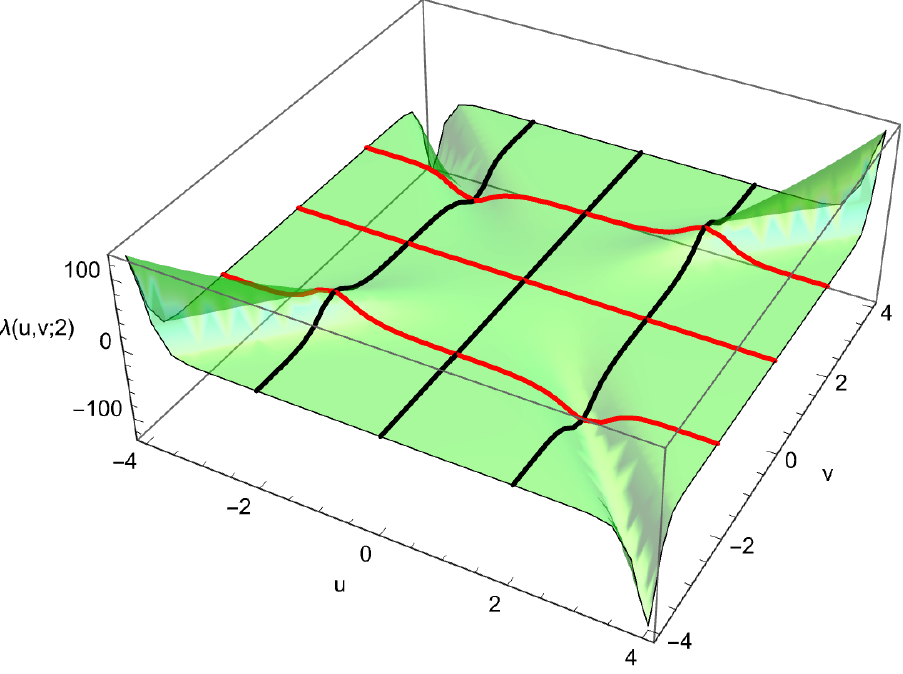}
\includegraphics[width=0.495\textwidth]{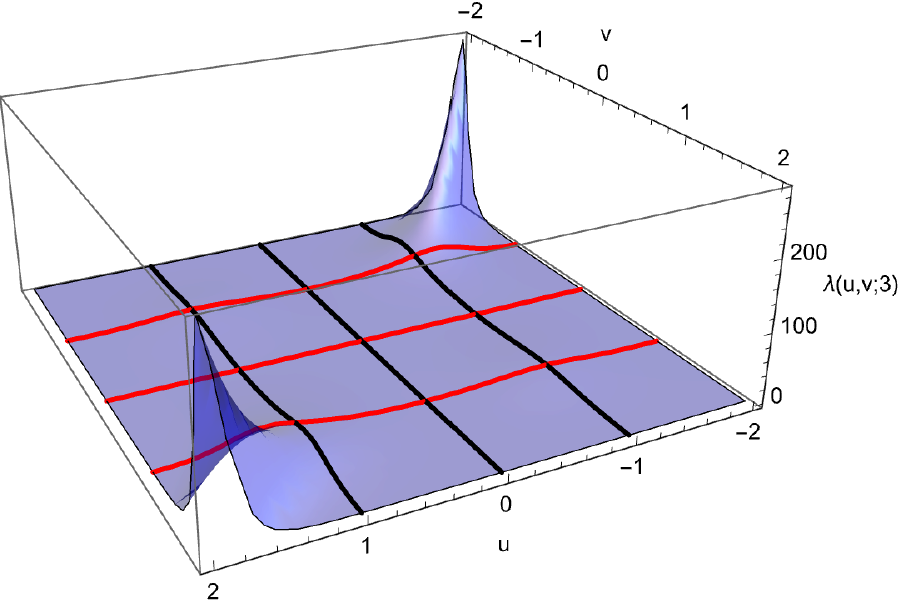}
\caption{The parametric solution from (\ref{eq11a}) with $k=1, ~\omega=1, ~\alpha=1/4$, for $p=1,2,3$. The traces  of each surface $u=-1/2,0,1/2 $ (black color),  and $v= -1/2,0,1/2$ (red color)
are also displayed.  These traces are the solitary waves  corresponding to  $p=1$,  $p=2$,  and $p=3$ from Fig. \ref{figabs}.}
\label{figa}
\end{figure}
%...................Fig. 3
\begin{figure}[ht!]
\centering
\includegraphics[width=0.49\textwidth]{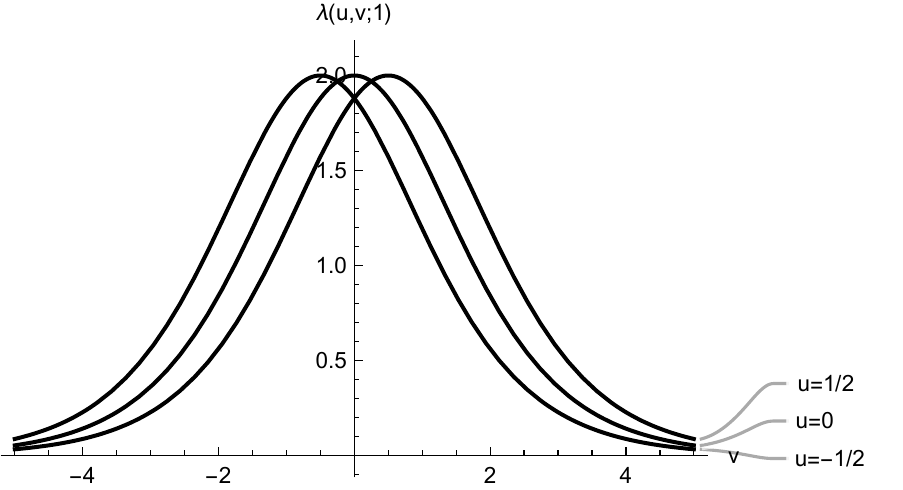}
\includegraphics[width=0.49\textwidth]{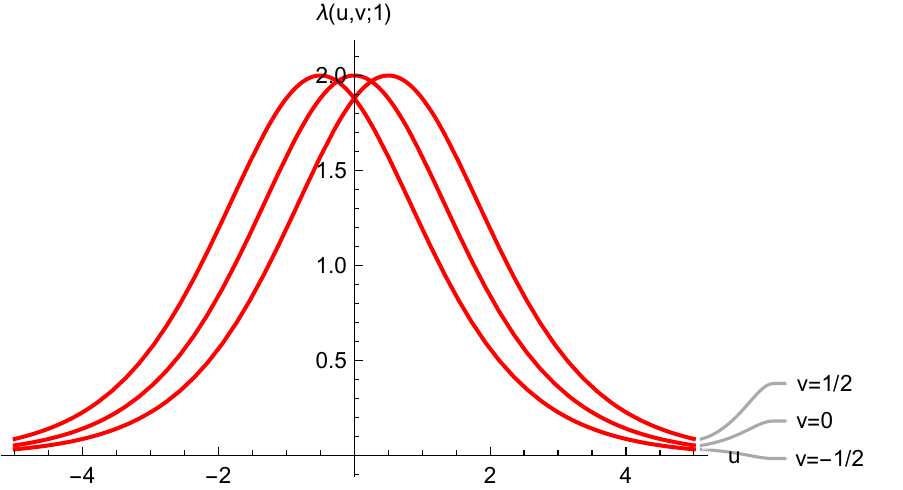}\\
\includegraphics[width=0.49\textwidth]{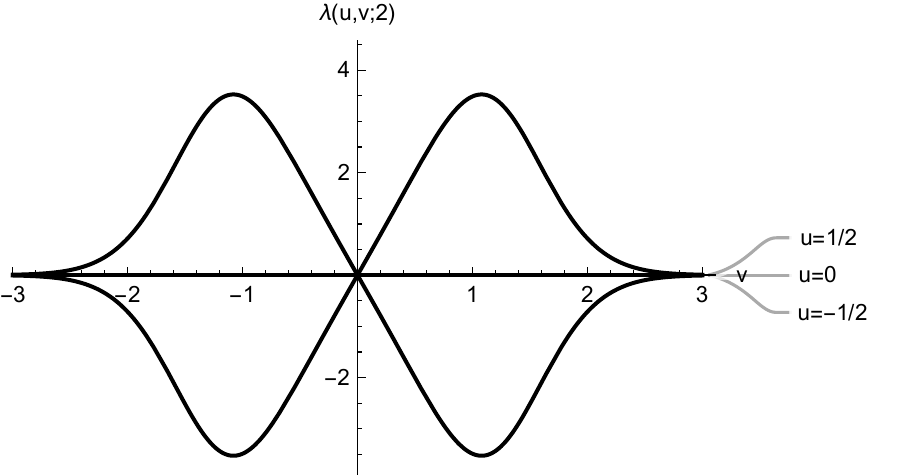}
\includegraphics[width=0.49\textwidth]{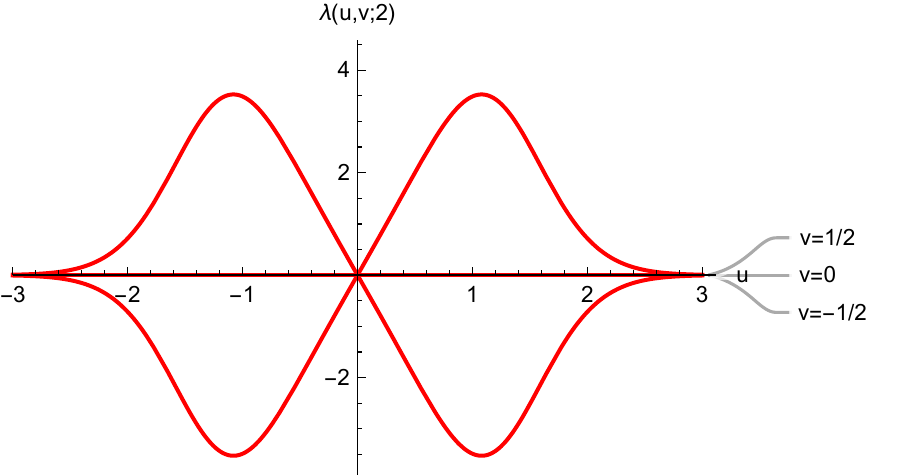}\\
\includegraphics[width=0.49\textwidth]{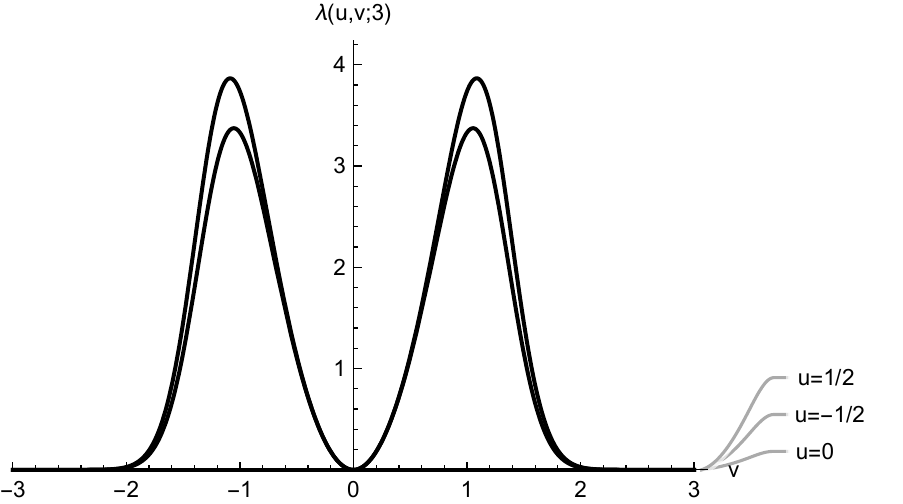}
\includegraphics[width=0.49\textwidth]{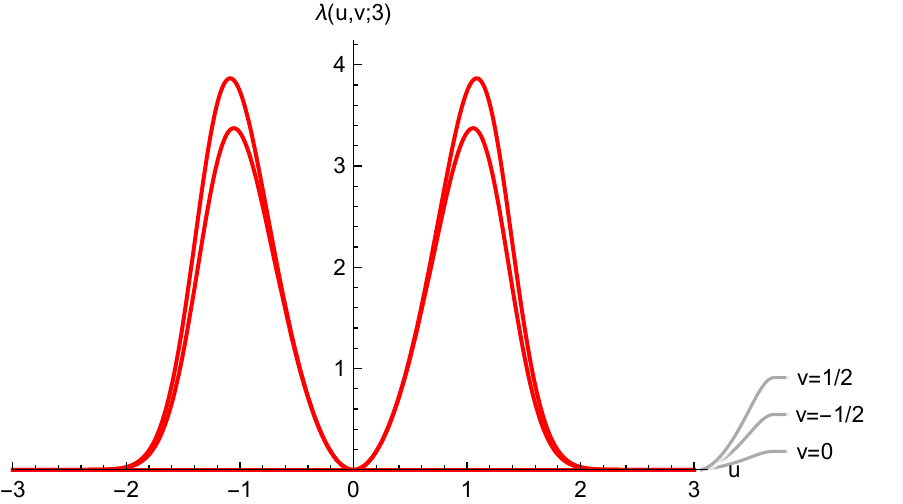}\\
\caption{Solitary waves for $p=1$ (top), $p=2$ (middle), and $p=3$ (bottom), corresponding to the red and black curves of Fig. \ref{figa}. The parameter $p$ acts a nonlinear power of the traveling wave variable,  but more importantly  as a parity operator of the amplitude of these pulses.  If $p$ is odd and solitons are even we have only bright solitons, while if $p$ is even we have odd solitons which are bright and dark.}
\label{figabs}
\end{figure}

\section{The modified variation of parameters method}

Next, we  present an equivalent method to obtain  \eqref{eq11} based on a traveling wave {\it ansatz}, and by using
a modified variation of parameters method which was first introduced by Ke\u{c}ki\'c in 1976 \cite{Keckic}.

Effecting the logarithmic derivative,  \eqref{eq2} becomes
%% eq. 28
\begin{equation}\label{eq4}
\lambda \lambda_{uv}-\lambda_u \lambda_v=\alpha {\lambda}^3~,
\end{equation}
which can be turned into the ordinary differential equation
%% eq. 29
\begin{equation}\label{eq5}
{\lambda _\xi}^2-\lambda \lambda _{\xi\xi}=\frac{\alpha}{k \omega}{\lambda }^3~,
\end{equation}
using the traveling wave variable $\xi= k u-\omega v$.
This equation can be written as a system of first order equations
%% eq. 30
\begin{equation}\label{sys1}
\begin{array}{l}
\lambda _\xi=\eta \equiv M(\lambda, \eta)~,\\
\eta_\xi=\frac{1}{\lambda }\left(  {\eta}^2-\frac{\alpha}{k \omega}{\lambda }^3\right)\equiv N(\lambda, \eta)~.
\end{array}
\end{equation}
The system \eqref{sys1} is integrable since there is a first integral ${\cal H}(\lambda,\eta)\equiv const$ of \eqref{eq5} such that
%% eq. 31
\begin{equation}\label{sys2}
	\begin{array}{l}
		M(\lambda, \eta)=\mu(\lambda,  \eta) \frac{\partial {\cal H}}{\partial  \eta}~,\\
		N(\lambda, \eta )=-\mu(\lambda,  \eta) \frac{\partial {\cal H}}{\partial \lambda}~.
	\end{array}
\end{equation}
For the special case of unitary integrating factor $\mu(\lambda, \eta)  \equiv 1$, then \eqref{sys1} is also Hamiltonian with first integral given by
%\begin{lemma}
%\end{lemma}
%\begin{proof}
%% eq. 32
\begin{equation}\label{halb}
			{\cal H}(\lambda, \eta)=\left(\frac{\eta}{\lambda}\right)^2+\frac{2 \alpha}{k \omega }\lambda \equiv const~,
\end{equation}
and Hamiltonian curves shown in Fig.~\ref{figham}. These curves are very important in the phase space ($\lambda,\lambda_\xi$) because they depict the first integrals given by the constant energy ${\cal H}=const$.
%..................FIGURE 4
\begin{figure}[ht!]
\centering
\includegraphics[]{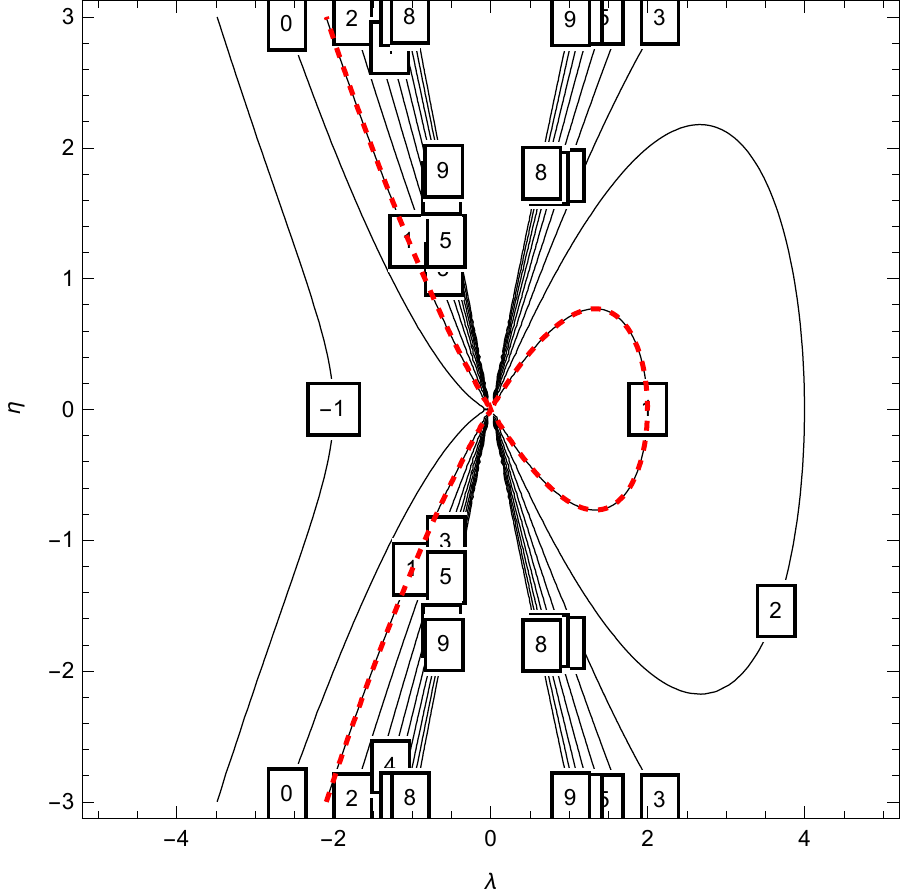}\\
\caption{Hamiltonian curves for  $k=1, ~\omega=1,$  and $\alpha=1/4$.  The dotted red curve corresponds to ${\cal H}=1$ and solution (\ref{eq88}).}
\label{figham}
\end{figure}
%.....................
However, our  system \eqref{sys2} holds true when  the integrating factor is
  $\mu(\lambda, \eta)=\lambda^2/2$  for which %$\square$
%\end{proof}
 \eqref{eq5} yields to the degenerate elliptic equation
 %% eq. 33
 \begin{equation}\label{hal}
 	{\lambda_\xi}^2={\cal H} {\lambda}^2-\frac{2\alpha}{k \omega}{\lambda}^3~.
 \end{equation}

When $a\rightarrow \infty$ then $\alpha \rightarrow 0$, and the   associated equation of \eqref{eq5} becomes
%% eq. 34
\begin{equation}\label{eq12}
{\lambda _\xi}^2-\lambda \lambda _{\xi\xi}=0~.
\end{equation}
By one quadrature we obtain
%% eq. 35
\begin{equation}\label{eq13}
\lambda_\xi=K \lambda~,
\end{equation}
while the second quadrature yields to the  exponential solution
%% eq. 36
\begin{equation} \label {eq14}
\lambda(\xi) = K_0 e^{K \xi}~,
 \end{equation}
with $K$ and $K_0$ arbitrary constants.

If we now suppose that $K$ is not a constant, i.e.,  $K= K(\lambda)$, by substituting (\ref{eq14}) into \eqref{eq5} we obtain
%% eq. 37
\begin{equation}\label{eq15}
{\lambda_\xi}^2-\lambda \lambda_{\xi\xi} = - K K_{\lambda} {\lambda}^3~,
\end{equation}
which can be matched with (\ref{eq5}) if $K(\lambda)$ is such that,
%% eq. 38
\begin{equation}
-K K_{\lambda}   = \frac{ \alpha}{k \omega}~.
\end{equation}
The latter equation leads to
%% eq. 39
\begin{equation}\label{we}
K(\lambda) = \sqrt{a_2 - \frac{2\alpha }{k\omega} \lambda}~,
\end{equation}
where $a_2$ is an arbitrary constant. Letting $a_3=-\frac{2\alpha}{k\omega}$  then (\ref{we}) becomes
%% eq. 40
\begin{equation} \label{eq19}
\lambda_{\xi} = \lambda\sqrt{a_2 +a_3 \lambda}~,
\end{equation}
which is exactly (\ref{hal}) for $a_2={\cal H}$. This equation belongs to the  general class of elliptic equations of the form (see also section 4 in \cite{gpr14})
%% eq. 41
\begin{equation}\label{eq2b}
{\lambda_\xi}^2 = \sum_{i=0}^3 a_i \lambda(\xi)^i \equiv Q_3(\lambda)~,
\end{equation}
which admit in general both cnoidal and solitary waves solutions. In our case, since $a_0 = a_1 = 0$,
then $\lambda=0$ is a double root of $Q_3(\lambda)$, and (\ref{eq2b}) admits
%This condition is achieved by choosing zero boundary conditions, i.e.,
%$h, h_{\xi}, h_{2\xi}\rightarrow 0$ as $|\xi|\rightarrow \infty$ which in turn yield to  in (\ref{eq6})
only the solitary wave solutions
%% eq. 41
\begin{equation}\label{eq8}
\lambda(\xi)=
- \frac{a_2}{a_3} \mathrm{sech}^2\left[\frac{1}{2} \sqrt{a_2}(\xi - \xi_0)\right], \quad a_2 > 0~,
\end{equation}
with $\xi_0$ an arbitrary constant.  For  $a_2=4 \alpha$, and by taking $\xi_0=0$ with  $a_3=-\frac{2\alpha}{k\omega}$,  this solution corresponds to   (\ref{eq11}) with integral of motion given by ${\cal H}=1$, and
%% eq. 43
 \begin{equation}\label{hal2}
 	{\lambda_\xi}^2= {\lambda}^2-\frac{{\lambda}^3}{2}~.
 \end{equation}
In particular,  for $\xi_0=0$,  we obtain the soliton
%% eq. 44
\begin{equation}\label{eq88}
\lambda(\xi)=2k\omega~ \mathrm{sech}^2\left(\frac{\xi }{2}  \right)~.
\end{equation}
To end up this section, we notice that according to the above discussion the centroaffine invariant can be written directly in terms of the Hamiltonian constant of the soliton solution as
%% eq. 45
\begin{equation}\label{dr}
{\cal I}=\left(\frac{{\cal H}}{k^2}\right)^3~,
\end{equation}
since for such solutions the Hamiltonian is a disguised form of the parameter entering the Liouville equation \eqref{eq1}.

\section{A one-parameter Lax Pair}

The second parametric  third order matrix Lax pair for \eqref{eq1}, known by Mikhailov \cite{Mikh}, is  given by
%% eq. 46
\begin{equation} \label{lax}
M = \begin{pmatrix} L_{11}\Lambda_u& 0 &L_{13}~e^{ \Lambda/2} \\ 0 & L_{22}\Lambda_u& 0\\ 0 &L_{32}e^{ \Lambda/2} & 0 \end{pmatrix}, \,\,\,\,\,\,\, N  = \begin{pmatrix} A_{11}\Lambda_v& 0 & 0 \\ 0 & A_{22}\Lambda_v&  A_{23}~e^{ \Lambda/2} \\A_{31}~e^{ \Lambda/2} &0 & 0 \end{pmatrix}~.
 \end{equation}
These matrices also satisfy
%% eq. 47
\begin{equation}\label{sys4}
\begin{array}{l}
R_u=MR~,\\
R_v=NR~,
\end{array}
\end{equation}
for some general vector wave function $R^T(u,v) =\left(F(u,v), G(u,v), H(u,v)\right)$.
%$R(u,v) = \begin{pmatrix} F(u,v) \\ G(u,v) \\ H(u,v) \end{pmatrix}$.
This time, the compatibility condition  leads to
%% eq. 48
  \begin{equation}\label{sy1}
\begin{pmatrix} A_{31} L_{13}\lambda - \frac{(A_{11}-L_{11}) (\lambda_u \lambda_v- \lambda \lambda_{uv})}{\lambda^2} & 0 & \frac{(1-2A_{11}) L_{13} \lambda_v}{2\sqrt{\lambda}}\\
0 & -A_{23}L_{32}\lambda + \frac{ (A_{22}-L_{22})(\lambda_u \lambda_v- \lambda \lambda_{uv})}{\lambda^2}  & \frac{A_{23}(L_{22}-1)}{2\sqrt{\lambda}}\lambda_u\\
\frac{-(1+2L_{11})A_{31}}{2\sqrt{\lambda}}\lambda_u  & \frac{(1+2A_{22})L_{32}}{2\sqrt{\lambda}}\lambda_v & (A_{23}L_{23}-A_{31}L_{13}) \lambda \end{pmatrix}\equiv 0~. \end{equation}
 Identifying the  terms in the matrix,  we require
 %% eq. 49
\begin{align} \label{eq400}
 \begin{array}{ll}
 & A_{11}=\frac{1}{2}~,\\
  & L_{11}=-\frac{1}{2}~, \\
 & L_{22}=\frac{1}{2}~, \\
  & A_{22}=-\frac{1}{2}~, \\
 & A_{23}L_{32}=A_{31}L_{13}~.
   \end{array}
 \end{align}
%  Using these conditions in the first two diagonal elements we obtain $A_{31}L_{13}=A_{23}L_{32}$.
Choosing the parameters $ L_{13}=\beta$ and $A_{31}=\frac{\alpha}{\beta} $, then $A_{23}= \beta $, and $B_{32}= \frac{\alpha}{\beta}$, and substituted into \eqref{lax} leads to the {\em one-parameter} Lax pair
 %% eq. 50
 \begin{equation} \label{lax2}
M = \begin{pmatrix} -\frac{\lambda_u}{2\lambda} & 0 &\beta \sqrt {\lambda}\\ 0 &  \frac{\lambda_u }{2\lambda} & 0\\ 0 &\frac{\alpha}{\beta}\sqrt{\lambda}  & 0 \end{pmatrix}, \,\,\,\,\,\,\, N  =
 \begin{pmatrix}
 \frac{\lambda_v}{2\lambda} & 0 & 0 \\ 0 & -\frac{\lambda_v}{2\lambda} &  \beta \sqrt{\lambda} \\ \frac{\alpha}{\beta}\sqrt{\lambda} &0 & 0
 \end{pmatrix}~.
 \end{equation}
% Employing this pair of matrices, \eqref{sy1} can be written in the standard form
% \begin{equation}
% L_v-A_u+[L,A]=\begin{pmatrix} -E(\Lambda) & 0 & 0\\ 0 & E(\Lambda)  & 0\\ 0 &0 & 0 \end{pmatrix} \equiv 0~.
% \end{equation}
 Now we will need to determine the eigenfunction $R(u,v)$ of the commutating differential operators which satisfy \eqref{sys4}.
 The first equation of the system leads to
 %% eq. 51
\begin{equation}\label{unu}
\begin{pmatrix} F_u + \frac{\lambda_u}{2\lambda}F-\beta \sqrt{\lambda}H \\ G_u-\frac{\lambda_u}{2\lambda}G \\ H_u-\frac{\alpha}{\beta}\sqrt{\lambda}G\end{pmatrix}=0~,
\end{equation}
while the second one gives
 %% eq. 52
\begin{equation}\label{doi}
\begin{pmatrix} F_v - \frac{\lambda_v}{2\lambda}F\\ G_v+ \frac{\lambda_v}{2\lambda}G -\beta\sqrt{\lambda} H \\ H_v-\frac{\alpha}{\beta}\sqrt{\lambda} F\end{pmatrix}=0~.
\end{equation}
To solve these systems, we notice that we can integrate the second equation of the first set \eqref{unu} to obtain
 %% eq. 53
\begin{equation}\label{z1}
G(u,v)=c_1(v) \text{sech}\left(\sqrt{\alpha }\xi\right)~. %(k u- \omega  v)\right]~.
\end{equation}
Using this result into the third equation  and integrating again we can find
 %% eq. 54
\begin{equation}\label{z2}
H(u,v)=\left[c_1(v)\frac{ \sqrt{2 \alpha \omega}}{\beta \sqrt{k}}\sinh \left(\sqrt{\alpha}\xi\right)+c_2(v)\cosh \left(\sqrt{\alpha}\xi\right)\right]
 \text{sech}\left(\sqrt{\alpha }\xi\right)~.
% (k u-\omega v )\right]+c_2(v)~.
\end{equation}
Since we have $H(u,v)$ up to the two integration constants, we integrate the first equation to obtain
 %% eq. 55
\begin{equation}\label{z3}
F(u,v)= \left\{\frac{\left[c_1(v)\sqrt{\alpha} \omega \sinh \left(\sqrt{\alpha}\xi\right)+  c_2(v)\sqrt{k \omega/2} \beta\cosh \left(\sqrt{\alpha}\xi\right)\right]^2}{\alpha  k \omega  c_1(v)}+c_3(v)\cosh^2 \left(\sqrt{\alpha}\xi\right)\right\}
\text{sech} \left(\sqrt{\alpha}\xi\right)~.
\end{equation}
The arbitrary constants $c_i$  will be determined from the second system \eqref{doi} by eliminating all the functions $F$, $H$ and their partial derivatives in $v$ to obtain  only a third order equation on $G$,  which after some cumbersome algebra reads
 %% eq. 56
\begin{equation}\label{mas}
G_{vvv}-3 \sqrt{\alpha} \omega  \tanh \left(\sqrt{\alpha}\xi\right) G_{vv} -(\sqrt{\alpha} \omega)^2 G_v+3 (\sqrt{\alpha} \omega)^3 \tanh \left(\sqrt{\alpha}\xi \right)G=0~.
\end{equation}
Once we obtained this equation,  we substitute $G(u,v)$ together  with its  partial derivatives,  $G_v$,  $G_{vv}$, and $G_{vvv}$ into \eqref{mas}, to hopefully obtain a manageable equation in $c_1$. Surprisingly,  but not unexpected, the new ODE in $c_1$  is
 %% eq. 57
 \begin{equation} \label{eq21a}
  c_1(v)''' - 4 \alpha \omega^2  c_1(v)'=  0~,
  \end{equation}
with the same equation  as \eqref{eq21} but with $k$ replaced by $\omega$, and $u$ by $v$.
Using
 %% eq. 58
\begin{align} \label{card3a}
 \begin{array}{ll}
  &c_1^1(v)= 1~,\\
 & c_1^2(v)= \sinh (2\sqrt{\alpha}\omega v)~,\\
 & c_1^3(v) = \cosh (2\sqrt{\alpha}\omega v)~,
\end{array}
 \end{align}
in \eqref{z1}, we obtain the solutions corresponding to $G(u,v)$ that are
 %% eq. 59
\begin{align} \label{card4a}
\begin{array}{ll}
&g^1(u,v)=\mathrm{sech} \left( \sqrt{\alpha}\xi\right)~,\\
&g^2(u,v)=  \sinh (2\sqrt{\alpha}\omega v)~\mathrm{sech} \left(\sqrt{\alpha}\xi\right)~, \\
&g^3(u,v)=  \cosh (2\sqrt{\alpha}\omega v)~\mathrm{sech} \left(\sqrt{\alpha}\xi\right)~.\\
\end{array}
\end{align}
The three solutions corresponding to $H(u,v)$ found from the second equation of  \eqref{doi} are
 %% eq. 60
\begin{align} \label{card4b}
\begin{array}{ll}
&h^1(u,v)= \sinh \left (\sqrt{\alpha} \xi\right)\mathrm{sech}  \left(\sqrt{\alpha}\xi\right)~,\\
&h^2(u,v)= \cosh  \left (\sqrt{\alpha} \zeta\right)~\mathrm{sech}  \left( \sqrt{\alpha}\xi\right)~, \\
&h^3(u,v)=    \mathrm{sinh}  \left(\sqrt{\alpha}\zeta\right)~\mathrm{sech}  \left(\sqrt{\alpha}\xi\right)~.\\
\end{array}
\end{align}
Finally,  since we have $H$, we can use the third equation of  system \eqref{doi} to find  the three corresponding $F(u,v)$, which are
 %% eq. 61
\begin{align} \label{card4c}
\begin{array}{ll}
&f^1(u,v)=\mathrm{sech} \left(\sqrt{\alpha}\xi\right)~,\\
&f^2(u,v)=  \sinh (2\sqrt{\alpha} k u)~\mathrm{sech} \left( \sqrt{\alpha}\xi\right)~, \\
&f^3(u,v)=  \cosh (2\sqrt{\alpha} k u)~\mathrm{sech} \left( \sqrt{\alpha}\xi\right)~.\\
\end{array}
\end{align}
Notice that these three functions satisfy the first equation of system \eqref{doi}, as they should.  Combining all these functions \eqref{card4a} - \eqref{card4c}, we can finally construct the two Darboux-transformed Liouville surfaces $\Sigma_2 (f^2,g^2,h^2)$, and $\Sigma_3 (f^3,g^3,h^3)$, which  are presented in Fig.~\ref{livi}.
%%... Fig. 5
\begin{figure}[ht!]
	\centering
	\includegraphics[width=0.495\textwidth]{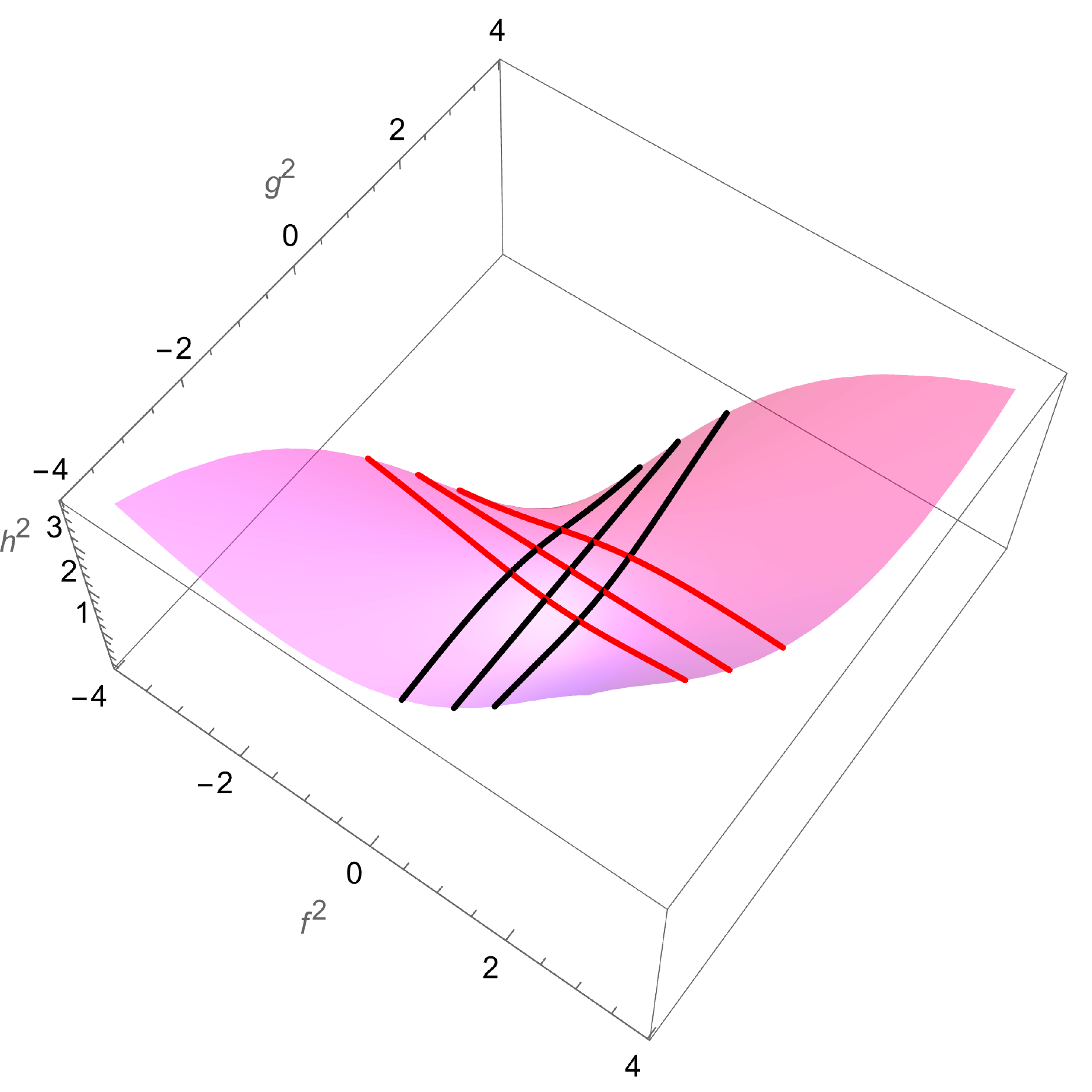}
	\includegraphics[width=0.495\textwidth]{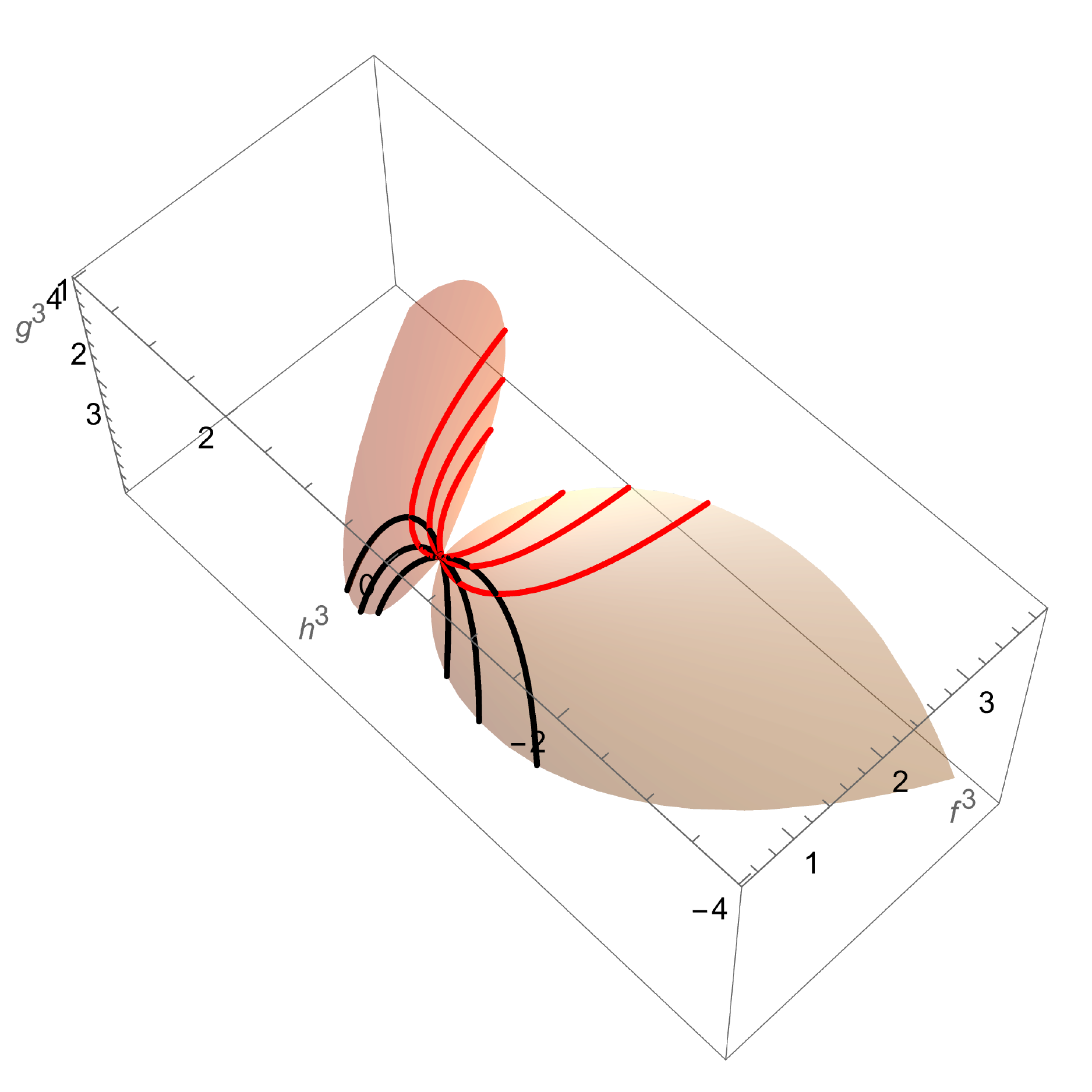}
	\caption{The Liouville soliton surfaces $\Sigma_{2}$ and $\Sigma_{3}$ corresponding to the Darboux-transformed Liouville solutions  \eqref{np2} for the same values  $k=1, \omega=1, \alpha=1/4$.  These are hyperbolic transformed  surfaces  corresponding to  systems \eqref{card4a}-\eqref{card4c}.  The traces  $u=-1/2,0,1/2$ (black color),  and $v= -1/2,0,1/2$ (red color) have now kink-like shapes.}
	\label{livi}
\end{figure}

\section{Summary and future prospects}

In this paper, we have shown explicitly how soliton surfaces can be constructed based on the soliton solution of the Liouville equation
in the travelling variable and its Darboux partner solutions. Similar surfaces can be constructed for other equations in the same class, such as the Tzitz\'eica
equation (also known as Dodd-Bullough-Tzit\'eica equation), the Dodd-Bullough-Mikhailov (DBM) equation, and the Tzitz\'eica-Dodd-Bullough-Mikhailov(TDBM) equation, which written as \eqref{eq5}, but with the coupling constant(s) in the power terms set to unity, have the form
 %% eq. 62
\begin{align}
		\begin{array}{ll}
			&{\lambda _\xi}^2-\lambda \lambda _{\xi\xi}=-\frac{1-{\lambda }^3}{k^2- \omega^2}~,\\
			&{\lambda _\xi}^2-\lambda \lambda _{\xi\xi}=-\frac{1+{\lambda }^3}{k\omega}~, \\
			&{\lambda _\xi}^2-\lambda \lambda _{\xi\xi}=-\frac{(1+\lambda){\lambda }^3}{k\omega}~, \\
		\end{array}
\end{align}
respectively. This set of equations have similar ${\rm sech}^2$ soliton solutions \cite{zna}, but the more complicated nonlinearities can generate cnoidal solutions. Consequently, periodic `cnoidal' surfaces can be constructed. In fact, for any cubic or quartic polynomial in the right hand side of nonlinear equations of this type one can construct surfaces associated to their solutions following the method described here for the monomic Liouville case. We plan to study these surfaces in a future publication.

\bigskip
\bigskip

\noindent {\bf Acknowledgment}\\

\noindent The authors wish to thanks the reviewers for useful comments and suggestions.

\bigskip
\bigskip

%\noindent {\bf Data availability statement}\\
%
%\noindent No new data were created or analysed in this study.
%
%The data cannot be made publicly available upon publication because they are not available in a format that is sufficiently accessible or reusable by other researchers. The data that support the findings of this study are available upon resonable request from the authors.
%
%\bigskip
%\bigskip
%
%\noindent {\bf Funding}\\
%
%This research did not receive any specific grant from funding agencies in the public, commercial, or not-for-profit sectors.
%
%\bigskip
%\bigskip
%
%\noindent {\bf Declaration of competing interests}\\
%
%The authors declare that they have no known competing financial interests or personal relationships
%that could have appeared to influence the work reported in this paper.

\bigskip
\bigskip

\noindent {\bf ORCID iDs}\\

\noindent S.C. Mancas http://orcid.org/0000-0003-1175-6869\\
\noindent K.R. Acharya http://orcid.org/0000-0003-3551-7141\\
\noindent H.C. Rosu http://orcid.org/0000-0001-5909-1945

%
%\bigskip
%\bigskip
%
%\noindent {\bf Credit authorship contribution statement}\\
%
%\noindent {\bf S.C.~Mancas}: Methodology, Writing – original draft.
%
%\noindent {\bf K.R.~Acharya}: Supervision, Validation.
%
%\noindent {\bf H.C.~Rosu}: Formal analysis, Project administration.


\begin{thebibliography} {99}

 \bibitem{sym83} Sym A 1983 Soliton surfaces II {\em Lett. Nuovo Cimento} {\bf 36} 307  %307-312

\bibitem{sym85} Sym A 1985 Soliton surfaces and their applications  %in Geometrical Aspects of the Einstein Equations and Integrable Systems
{\em Lecture Notes in Physics} {\bf 239} editor Martini R (Berlin: Springer) 154   %154-231
	

\bibitem{cies98} Cie\'sli\'nski J 1998 The Darboux-Bianchi-B\"acklund transformation and soliton surfaces {\em Proceedings of The First Non-Orthodox School on
Nonlinearity and Geometry} editors Wojcik D and Cie\'sli\'nski J (Warszawa: Polish Scientific Publishers PWN) arXiv:1303.5472

\bibitem{gp12} Grundland A M, Post S 2012 Soliton surfaces via a zero-curvature representation of differential equations
{\em J. Phys. A: Math. Theor.} {\bf 45} 115204

\bibitem{gpr14} Grundland A M, Post S, Riglioni D 2014 Soliton surfaces and generalized symmetries of integrable systems
{\em J. Phys. A: Math. Theor.} {\bf 47} 015201

\bibitem{perel87} Perelomov A M 1987 Chiral models: Geometrical aspects {\em Phys. Rep.} {\bf 146} 135  %135-213

\bibitem{dt07} Dai B, Terng C-L 2007 B\"acklund transformations, Ward solitons, and unitons {\em J. Diff. Geom.} {\bf 75} 57    %57-108 (2007).

\bibitem{Mikh} Mikha\u{\i}lov A V 1979 Integrability of a two-dimensional generalization of the Toda chain {\em JETP Lett.} {\bf 30} 414  %414-418 (1979).
%30 Years Of The Landau Institute Selected Papers. 1996; 377-381. 10.1142/9789814317344_0050

  \bibitem{mo} Monge G 1850 {\em Application de l'Analyse \`a la G\'eom\'etrie} 5th Edition (Paris: Bachelier) 597

\bibitem{Lio} Liouville J 1853 Sur l'\'equation aux diff\'erences partielles $d^2\,log\lambda/dudv\pm \lambda/2a^2=0$ {\em J Math. Pures Appl.} %Serie 1.
 {\bf 18} 71  %{\bf 18} 71-72 (1853)

 \bibitem{bn} Barbashov B M, Nesterenko V V 1980 Differential geometry and nonlinear field models {\em Fortschritte der Physik} {\bf 28} 427  %-464 (1980)

 \bibitem{teschner} Teschner J 2001, Liouville theory revisited {\em Class. Quant. Grav.} {\bf 18}, R153

 \bibitem{dpp} Dzordzhadze G P, Pogrebkov A K, Polivanov M K 1979 Singular solutions of the equations
$\square \phi+(m^2/2)\exp \phi=0$ and dynamics of singularities {\em Theor. Math. Phys.} {\bf 40} 221

\bibitem{jkm} Johansson L, Kihlberg A, Marnelius R 1984 Sectors of solutions and minimal energies in classical Liouville theories for strings
{\em Phys. Rev. D} {\bf 29}, 2798  %-2813 (1984)

\bibitem{TG} Tzitz\'eica G 1924 {\em G\'eom\'etrie Diff\'erentielle Projective des R\'eseaux} (Bucharest: Cultura Na\c{t}ional\u{a})

\bibitem{TG1} Tzitz\'eica G 1910 Sur une nouvelle classe de surfaces {\em Comptes Rendus Acad. Sci. Paris} {\bf 150} No. 16 955    %955-956 (1910).

\bibitem{TG2} Tzitz\'eica G 1910 Sur une nouvelle classe de surfaces {\em Comptes Rendus Acad. Sci. Paris} {\bf 150} No. 20 1227   %1227-1229 (1910).

\bibitem{Bre} Brezhnev Yu V 1996 Darboux transformation and some multi-phase solutions of the Dodd-Bullough-Tzitz\'eica equation: $U_{xt}=e^U-e^{-2U}$
{\em Phys. Lett. A} {\bf 211} 94   %94-100 (1996).

\bibitem{Schief} Schief W K, Rogers C 1994 The affinsph\"aren equation. Moutard and B\"acklund transformations {\em Inverse Problems} {\bf 10} 711    %711-731 (1994).

 \bibitem{Keckic} Ke\u{c}ki\'c J D 1976
 Addition to Kamke's Treatise, VII: Variation of parameters for nonlinear second order differential equations
 {\em Univ Beograd. Publ Elektrotehn FAK Ser. Mat. Fis. No. 544 - No. 576} 31   %31-36 (1946).

\bibitem{zna} Mancas S C, Rosu H C, P\'erez-Maldonado M 2018 Traveling-wave solutions for wave equations with two exponential nonlinearities
{\em Zeitschrift f\"ur Naturforschung A} {\bf 73} 883   %883-892 (2018).

%A. S. Fokas, I. M. Gel'fand, F. Finkel* and Q. M. Liu, A formula for constructing infitely many surfaces
%on Lie algebras and integrable equations, Sel. math., New ser. 6 (2000) 347-375.
%LUIGI BIANCHI
%Ricerche sulle superficie elicoidali e sulle superficie a curvatura costante
%Annali della Scuola Normale Superiore di Pisa, Classe di Scienze 1re série, tome 2 (1879), p. 285-341


\end{thebibliography}
\end{document}